\documentclass[prb,twocolumn]{revtex4}
\usepackage{bm}
\usepackage{graphicx}
\usepackage{amsmath}
\usepackage{eufrak}
\usepackage{color}
\begin{document}

\title{Ratchet effects induced by terahertz radiation in heterostructures\\ with a lateral
periodic potential}

\author{P. Olbrich$^1$, E.L. Ivchenko$^2$, T. Feil$^1$, R. Ravash$^1$, S.D. Danilov$^1$, J.
Allerdings$^1$, D. Weiss$^1$, and
S.\,D.~Ganichev$^{1}\footnote{e-mail:
sergey.ganichev@physik.uni-regensburg.de}$}
\affiliation{$^1$  Terahertz Center, University of Regensburg,
93040 Regensburg, Germany}
\affiliation{$^2$A.F.~Ioffe Physico-Technical Institute, Russian
Academy of Sciences, 194021 St.~Petersburg, Russia}

\begin{abstract}
We report on the observation of terahertz radiation induced
photogalvanic currents in semiconductor heterostructures with
one-dimensional lateral periodic potential. The potential is
produced by etching a grating into the sample surface. The
electric current response is well described by phenomenological
theory including both the circular and linear photogalvanic
effects. Experimental data demonstrate that the inversion
asymmetry of the periodic lateral pattern can be varied by means
of electron beam lithography to produce classical lateral
ratchets. A novel microscopical mechanism for the
polarization-dependent photogalvanic effects has been proposed to
interpret the experimental findings. The photocurrent generation
is based on the combined action of the lateral periodic potential
and the modulated in-plane pumping. The latter modulation stems
from near-field effects of the radiation propagating through the
grating.

\end{abstract}


\maketitle
\section{Introduction}
Nonequilibrium spatially-periodic noncentrosymmetric systems are
able to transport particles in the absence of an average
macroscopic force. The directed transport in such systems,
generally known as the ratchet effect, has a long history and is
relevant for different fields of
physics~\cite{reimann,applphys,nori,kotthaus,grifoni,science,samuelson,grifoni2}.
If this effect is induced by electro-magnetic radiation it is
usually referred to as photogalvanic (or photovoltaic) effect,
particularly if breaking of the spatial inversion symmetry is
related to the microscopic structure of the
system~\cite{1,2,chepel,3,regensburg}. Blanter and
B\"uttiker~\cite{buttiker2,buttiker} have shown that one of the
possible realizations of the Seebeck ratchet can be a superlattice
irradiated by light through a mask of the same period but phase
shifted with respect to the superlattice. In the present work we
have experimentally realized this idea with some modifications.
The photocurrent has been observed in semiconductor
heterostructures with a one-dimensional lateral periodic potential
induced by etching a noncentrosymmetric grating into the sample
cap layer. The in-plane modulation of the pump radiation appears
hence not via a mask with periodic structures but due to the
near-field effects of radiation propagating through the grating.
This photothermal ratchet effect was predicted by
B\"uttiker~\cite{buttiker2} and is polarization
independent under normal light incidence. In addition we have
observed photocurrents sensitive to the plane of polarization of
the linearly polarized terahertz (THz) radiation and to the
helicity in case of circularly polarized photoexcitation. The
theoretical analysis enables us to propose new mechanisms of the
observed circular and linear photogalvanic effects which are
related to the combined action of out-of-phase periodic potential
and in-plane modulated pumping of the two-dimensional electron
system.

\section{Samples and experimental methods}

Here we study photocurrents in (001)-grown GaAs/AlGaAs
heterostructures with superimposed lateral grating having a period
of 2.5 $\mu$m. The electronic micrograph are shown in the inset of
Fig.~\ref{fig2}. The lateral gratings are prepared on
molecular-beam epitaxy (100)-grown Si-$\delta$-doped n-type
GaAs$/$Al$_{0.3}$Ga$_{0.7}$As quantum-well structures. The
mobility $\mu$ and carrier density $n_s$ measured at 4.2 K in our
single QW structure of 30~nm width are $\mu = 4.82 \times
10^6$~cm$^{2}$/Vs and $n_s = 1.7 \times 10^{11}$~cm$^{-2}$.
Samples grown along $z
\parallel [100]$ were square shaped with sample edges of 5~mm
length oriented along $[1{\bar 1}0]$ and $[110]$. To measure
photocurrents, pairs of ohmic contacts were alloyed in the middle
of sample edges. Gratings of 0.5 $\mu$m width and period 2.5
$\mu$m are obtained by electron beam lithography and subsequent
reactive ion etching using SiCl$_4$. Care was taken not to etch
through the two-dimensional electron gas. To get a large patterned
area of about 1.4 mm$^2$, 64 squares, each  150$\mu$m $\times$ 150 $\mu$m,
were stitched together. The one-dimensional gratings are oriented
either along $\left\langle 010 \right\rangle$ (sample A) or close
to $\left\langle 110\right\rangle$ (samples B and C)
crystallographic directions. In the latter case the grating is
slightly misaligned by a small angle of about 4$^{\circ}$ with
respect to the crystallographic direction. While the cross section
of grooves prepared close to the $\left\langle 110\right\rangle$
crystallographic direction is rather symmetric the shape of
grooves prepared along $\left\langle 010 \right\rangle$
crystallographic direction is substantially asymmetric. The
average depth on the right side of the groove is smaller than that
on the left side. The reason for this might be attributed to the
difference in the etching velocities along [110] and [1$\bar{1}$0]
directions~\cite{adachi,adachi2}.

\begin{figure}[t]
\includegraphics[width=0.85\linewidth]{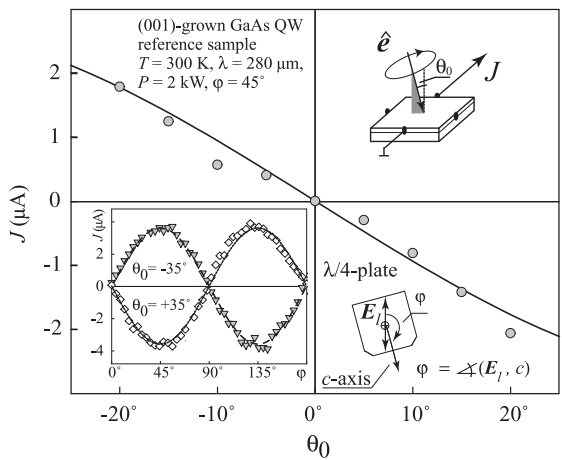}
\caption{Circular photogalvanic current $J_{C} = [J(\varphi = 45^\circ) -
J(\varphi = 135^\circ)]/2$ as a function of angle of incidence
$\theta_0$ measured in a GaAs$/$Al$_{0.3}$Ga$_{0.7}$. As reference QW sample without
lateral structure. The current  is measured in the direction normal to  light
propagation. Photocurrent is excited by radiation with wavelength
$\lambda$~=~280~$\mu$m and power $P \approx 2$~kW. The inset
(bottom, left) shows the dependence of the total photocurrent $J$ on 
angle $\varphi$ measured for angles of incidence $\theta_0 = \pm 35^\circ$. Two
other insets (right panels) show,
respectively, the experimental geometry and the quarter-wave plate
which varies the radiation helicity according to $P_{\rm circ}=
\sin 2 \varphi$. Full lines are fits to the phenomenological theory
for $C_{2v}$ symmetry relevant for (001)-grown unstructured III-V
QWs and given by Eq.~\protect(\ref{jref}), see~\protect\cite{10}. } \label{fig1}
\end{figure}

\begin{figure}[t]
\includegraphics[width=0.95\linewidth]{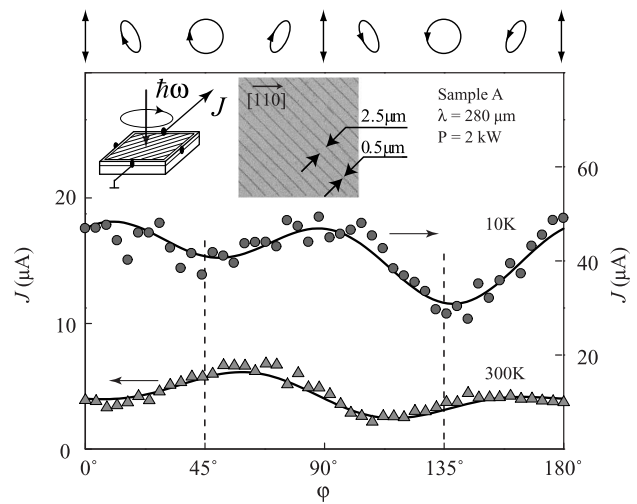}
\caption{Photocurrent measured as a function of the angle $\varphi$ at
normal incidence ($\theta_0 = 0^\circ$) in sample A with the
asymmetric lateral structure prepared along the [010] crystallographic
axis. The current is measured at room
temperature and T = 10~K, excited by the radiation
with wavelength $\lambda$~=~280~$\mu$m and power $P \approx 2$~kW.
Full line are fits to Eq.~\protect(\ref{phenom1}) (see also Eq.~\protect(\ref{3phi})). Left inset
shows the experimental geometry, and central inset
displays a micrograph of the grating. The ellipses on top illustrate the state of polarization for various angles
$\varphi$.} \label{fig2}
\end{figure}

For optical excitation we used pulsed molecular THz lasers with
NH$_{3}$ as an active medium \cite{book}. Circularly and linearly
polarized radiation pulses of about 100~ns duration with
wavelength $\lambda$= 280 $\mu$m and power $P \simeq $ 2~kW were
applied. The photocurrents were induced by indirect intrasubband
(Drude-like) optical transitions in the lowest size-quantized
subband. To measure polarization dependencies we used $\lambda$/4
plates for conversion of linear to circular polarization. The
helicity is described by $P_{\rm circ}= \sin 2 \varphi$, where
$\varphi$ is the angle between the initial plane of laser beam
polarization $\bm{E}_l$ and the $c$-axis of the $\lambda$/4 plate.
To investigate the photogalvanic effects we also used linearly
polarized light. In the experiments the plane of polarization of
the radiation, incident on the sample, was rotated by
$\lambda/2$ plates. This enabled us to vary the azimuth angle
$\alpha$ from $0^\circ$ to $180^\circ$ covering all possible
orientations of the electric field vector in the interface plane.
Radiation was applied at oblique incidence described by the angle
of incidence $\theta_0$ varying from $-25^\circ$ to +25$^\circ$
(Fig.~\ref{fig1}) and at normal incidence (Fig.~\ref{fig2}).
The current generated by THz-light in the unbiased samples was
measured via the voltage drop across a 50~$\Omega$ load resistor
in a closed-circuit configuration. The voltage was recorded with a
storage oscilloscope.

In (001)-oriented unpatterned samples a signal is only detectable
under oblique incidence. The photocurrent measured perpendicularly
to the wave vector of the incident light is almost proportional to
the helicity $P_{c}$ and reverses its direction when the
polarization switches from left-handed to right-handed circular
(see the inset panel of Fig.~\ref{fig1}). A photocurrent, but of
substantially smaller magnitude, is also generated by applying
linearly polarized radiation. In the whole temperature range from
room temperature to 4.2 K and for  excitation with both circularly
as well as linearly polarized radiation the variation of the angle
of incidence from $\theta_0$ to $-\theta_0$ changes the sign of
the photocurrent $J$. This is shown in Fig.~\ref{fig1} for the
circular photogalvanic effect (CPGE) obtained after
$J_{C} = [J(\varphi = 45^\circ) - J(\varphi = 135^\circ)]/2$. For
normal incidence the photocurrent vanishes. The photocurrent is
well described by the phenomenological theory of the circular and
linear photogalvanic effect obtained for the point group C$_{2v}$
which is relevant for this type of structures~\cite{PRL01}. Theory
yields for the dominating CPGE photocurrent $J_{\rm ref}$  of the unpatterned
reference sample:
\begin{equation} \label{jref}
J_{\rm ref} = a_{\rm ref} \sin{\theta_0} \xi P_c \:,
\end{equation}
where $a_{\rm ref}$ is a constant, $\xi = t_pt_s/t_0^2$, $t_p$ and
$t_s$ are the Fresnel transmission coefficients for the $p$- and
$s$-polarized light, respectively, and $t_0$ is the transmission
coefficient under normal incidence. The corresponding fits of our
data are shown in  Fig.~\ref{fig1} by the full line and in the
inset in Fig.~\ref{fig1} as dashed and dotted lines.

\begin{figure}[t]
\includegraphics[width=0.8\linewidth]{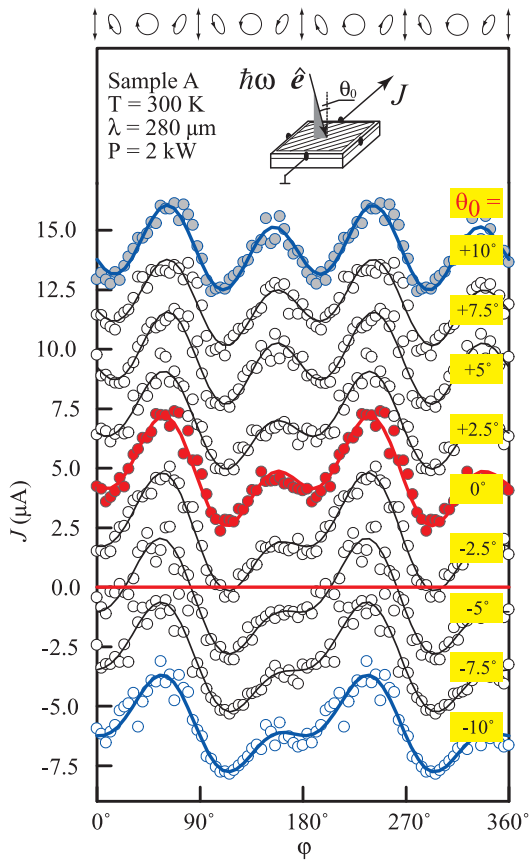}
\caption{Photocurrent measured as a function of the angle $\varphi$ at
various angles of incidence ($\theta_0$) in sample A with an
asymmetric lateral structure prepared along the [010] crystallographic
axis. The data for $\theta_0 \neq 0$ are shifted by $+ 2.5$~$\mu$A for each $2.5^{\circ}$ step
(positive $\theta_0$) and
by  $- 2.5$~$\mu$A for each $2.5^{\circ}$ step (negative $\theta_0$).
The current is measured at room
temperature, excited by  radiation
with wavelength $\lambda$~=~280~$\mu$m and power $P \approx 2$~kW.
Full lines are fits to Eqs.~\protect(\ref{phenom1}) (see also Eq.~\protect(\ref{3phi})). The inset shows the experimental geometry. 
The ellipses on top illustrate the state of polarization for various
angles $\varphi$. The data, periodic in $\pi$, are plotted from $\varphi = 0^{\circ}$ to $\varphi =
360^{\circ}$ to better visualize the features of interest.
} \label{fig2wf}
\end{figure}

\begin{figure}[t]
\includegraphics[width=0.85\linewidth]{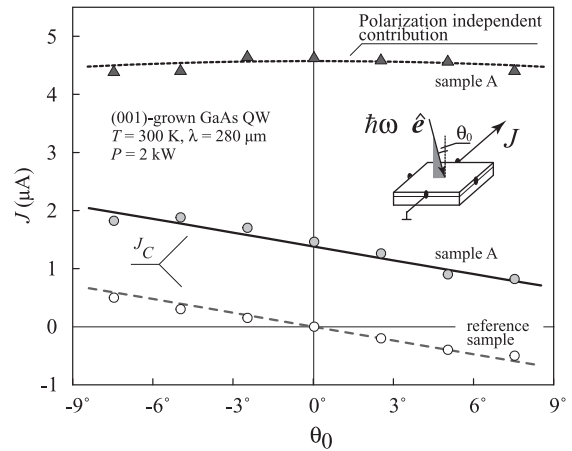}
\caption{Circular photogalvanic current $J_{C} = [J(\varphi = 45^\circ) -
J(\varphi = 135^\circ)]/2$ measured as a function of the incidence
angle $\theta_0$ in sample A with the asymmetric lateral structure
along [010] crystallographic axis (full circles) and the
unstructured reference QW sample (open circles). The current is
measured in the direction normal to light propagation and
excited by radiation with wavelength $\lambda$~=~280~$\mu$m and
power $P \approx 2$~kW. Full and dashed lines are fits to Eqs.~\protect(\ref{jrefcirc}), see
also the term proportional to $P_c$ in Eq.~\protect(\ref{3phi}), and \protect(\ref{jref}), respectively.
Triangles show the $\theta_0$ dependence of the
polarization-insensitive photocurrent together with the curve
calculated according to $J = d \xi \cos{\theta_0}$ following from
Eq.~\protect(\ref{1}). The data are
obtained from fitting the $\varphi$-dependence of the photocurrent
$J$ taken at various incidence angles. Inset shows the
experimental geometry used for sample A.
 } \label{fig3}
\end{figure}

The situation changes drastically for samples with grating. Now a
photocurrent is also detected at normal incidence. The width of the
photocurrent pulses is about 100 ns which corresponds to the THz
laser pulse duration.
In the patterned samples with the grooves oriented along
$\left\langle 010 \right\rangle$  we observed that the magnitude
of the photocurrent detected at normal incidence (Fig.~\ref{fig2})
is comparable and even larger than that obtained in the reference
sample at large angles of incidence (Fig.~\ref{fig1}). Moreover
the polarization behaviour has changed.
Figure~\ref{fig2} shows ellipticity dependent measurements of sample A excited at normal incidence.
The data can be well fitted by
\begin{equation} \label{phenom1}
J = a \sin 2\varphi + b \sin 4\varphi + c \cos 4\varphi + d\:.
\end{equation}
Here, the parameters $a$, $b$, $c$, and $d$ are phenomenological
fitting parameters, described below. Figure~\ref{fig2} shows that
the photon helicity dependent photocurrent caused by
circular photogalvanic effect gives an essential contribution.
Figure~\ref{fig3} displays the CPGE current as a function of the
angle of incidence $\theta_0$ for sample A (open circles) and the
unstructured reference sample (full circles). To extract the CPGE
current from the total current we used the fact that the CPGE
contribution, given by the term $a \sin 2\varphi$, changes its sign
upon switching the helicity while all the other terms remain
unchanged. Taking the difference of photocurrents of right and
left handed radiation we get the CPGE current
$J_{C}$. At oblique incidence, in the structured sample it
consists of two contributions. The first one has the same origin
as the one observed in the reference sample and is described by
Eq.~(\ref{jref}). The second one is due to the lateral structure.
The dependence of the CPGE photocurrent on the angle of incidence
$\theta_0$ can be well fitted by
\begin{equation} \label{jrefcirc}
J_{C} = ( a_{\rm ref} \sin{\theta_0} + a \cos{\theta_0}) \xi  \:.
\end{equation}

\begin{figure}[t]
\includegraphics[width=0.85\linewidth]{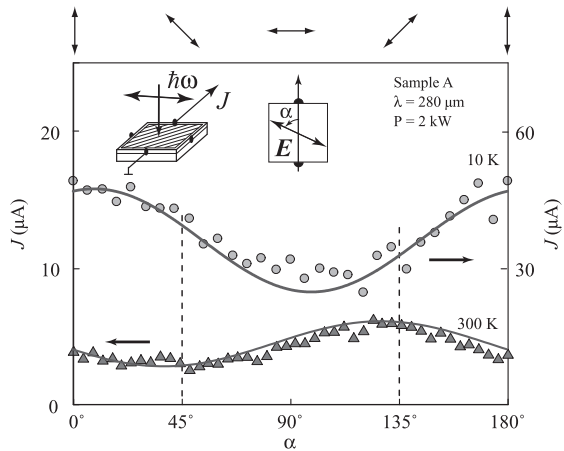}
\caption{Photocurrent $J$ measured as a function of the angle $\alpha$ under normal incidence
at room temperature and T = 10~K
in sample A with the asymmetric lateral
structure along [010] crystallographic axis.
Photocurrent is excited by linearly polarized radiation with
wavelength $\lambda$~=~280~$\mu$m and power $P \approx 2$~kW. Full
lines are fits to Eq.~\protect(\ref{phenom2}), see also Eq.~\protect(\ref{3alpha}). We used for fitting
the same values of $b$, $c$ and $d$ as in the
experiments with elliptically polarized radiation, see Fig.~\ref{fig2}. Left inset shows
the experimental geometry, and right inset
defines the angle $\alpha$. On top the arrows indicate the polarization corresponding to various values of
$\protect\alpha$. } \label{fig4}
\end{figure}

Now we turn to the photon helicity independent contributions to
the photocurrent, denoted by coefficients $b$, $c$ and $d$ in
Eq.~(\ref{phenom1}). These contributions we attribute to the
linear photogalvanic effect (LPGE). The LPGE
photocurrents can be generated by applying linearly polarized
radiation. Figure~\ref{fig4} shows the dependence of the
photocurrent on the azimuth angle describing the variation of the
light's electric field direction relative to the crystallographic
direction [110]. We found that all data can be well fitted by
\begin{equation} \label{phenom2}
J = 2b \sin 2\alpha + 2c \cos 2\alpha + d - c\:.
\end{equation}
We emphasize that $b$, $c$, and $d$ are the same fitting parameters used for the data shown
in Fig.~\ref{fig2}. 
Figure~\ref{fig3} shows the dependence of the polarization
independent contribution, proportional to the coefficient $d$, on
the  angle of incidence $\theta_0$. In this case the experimental
data can be well fitted by
\begin{equation} \label{jd}
J = d \cos{\theta_0} \xi    \:.
\end{equation}
Figures~\ref{fig2} and~\ref{fig4} demonstrate that the dominant
contribution to the photocurrent is polarization independent and
can therefore be obtained by unpolarized radiation. In  samples B
and C with the grooves oriented along $\left\langle
110\right\rangle$  crystallographic directions we also detected a
photocurrent at normal incidence having the same polarization
dependences as sample A, Eqs.~(\ref{phenom1}) and (\ref{phenom2}).
However,  the photocurrent measured in sample A is about an order
of magnitude larger than in sample B and C. We ascribe this to the
grooves profile being strongly asymmetric in sample A while nearly
symmetric in samples B and C.


%

%

\section{Phenomenological description}

In this section our theoretical analysis of the experimental data
is based on the symmetry considerations of the phenomenological
equations describing the photogalvanic effects (PGE) under study.
Under normal incidence of the laser radiation on the sample the
in-plane photocurrent  is given by
\begin{equation} \label{1}
j_l = I \sum\limits_{m,n} \chi_{lmn} \{ e_m e^*_n \} + I
\gamma_{lz} P_c\:,
\end{equation}
where  $l,m,n$ are the in-plane coordinates, $\{ e_m e^*_n \} = ( e_m e^*_n + e_n e^*_m)/2$, $I$, ${\bm
e}$ and $P_c$ are the light intensity, polarization unit vector
and degree of circular polarization. The third rank tensor
$\chi_{lmn}$ describing the linear PGE is symmetrical with respect
to the interchange of the second and third indices, and $\gamma_{lm}$
is a second-rank pseudotensor describing the circular PGE. The
(001)-grown heterostructure has the point-group symmetry $F_{\rm
ref} =$ C$_{2v}$ which forbids in-plane photocurrents under normal
incidence~\cite{book} as it is confirmed by measurements performed on
the reference samples.

\begin{figure}[t]
\includegraphics[width=0.8\linewidth]{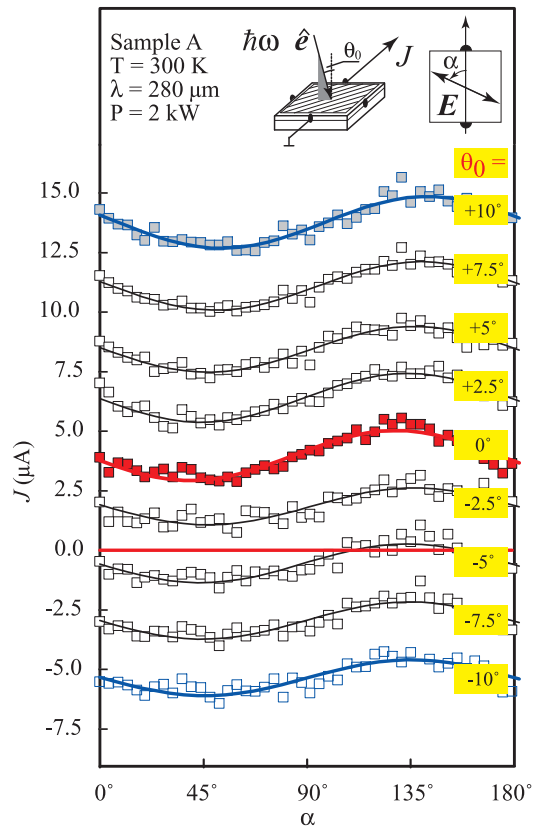}
\caption{ Photocurrent $J$ measured as a
function of the angle $\alpha$ at room temperature
in sample A with the asymmetric lateral
structure along [010] crystallographic axis.
The data for $\theta_0 \neq 0$ are shifted by $+ 2.5$~$\mu$A for each $2.5^{\circ}$ step (positive $\theta_0$) and
by  $- 2.5$~$\mu$A for each $2.5^{\circ}$ step (negative $\theta_0$).
The current is measured at room
temperature, excited by  radiation
with wavelength $\lambda$~=~280~$\mu$m and power $P \approx 2$~kW.
Full lines are fits to Eqs.~\protect(\ref{phenom1}), see also Eq.~\protect(\ref{3alpha}).  The left inset
shows the experimental geometry. The right inset 
displays the sample and the radiation electric field 
viewing from the source of radiation side.}  \label{fig2wfa}
\end{figure}

The lateral superlattice can reduce the symmetry of the system.
Let us denote the superimposed periodic lateral potential as
$V({\bm \rho})$, where ${\bm \rho}$ is the two-dimensional
radius-vector, and introduce the period $a$ and two in-plane axes
$x$ and $y$ oriented, respectively, parallel and perpendicular
to the direction of periodicity. Then, by definition, the lateral
superlattice potential does not depend on $y$ and is a periodic
function of $x$, namely, $ V(x + a) = V(x)$. One of the
symmetry elements of this potential is the mirror reflection plane
$\sigma_{y}$ perpendicular to the axis $y$. Its total
point-group symmetry $F_{\rm SL}$ can be either C$_s$ if the
function $V(x)$ is asymmetric or C$_{2v}$ if $V(x)$ is an even
function with respect to a certain origin on the axis $x$. In
the latter case, in addition to the identity element $e$ and the
mirror plane $\sigma_{y}$, the point group contains the mirror
plane $\sigma_{x} \perp x$ and the second-order rotation axis
$C_2 \parallel z$.

The symmetry of the structured sample is determined by the direct
product $F_{\rm ref} \times F_{\rm SL}$ of the point groups
describing the symmetries of the reference heterostructure and
lateral potential. If the potential $V(x)$ is symmetrical both
groups $F_{\rm ref}$ and $F_{\rm SL}$ have a common element $C_2$
and the photocurrents under normal incidence are forbidden. This
is obviously realized in the samples with the superlattice axes
$x, y$ oriented in the $\langle 110 \rangle$ directions. Thus, the
photocurrents can be induced under normal incidence only in the
case of an asymmetrical potential $V(x)$. If the axes $x, y$ of
this potential coincide with the axes [1$\bar{1}$0], [110], the
symmetry of the system is C$_s$, Eq.~(\ref{1}) reduces to
\begin{eqnarray} \label{2}
j_{x} &=& I [\chi_1 + \chi_2 (|e_{x}|^2 - |e_{y}|^2)]\:,\\
j_{y} &=& 2 I \chi_3  \{ e_{x} e^*_{y} \} + I \gamma P_c\:,\nonumber
\end{eqnarray}
and is governed by four linearly independent coefficients. If the
axes $x, y$ are rotated with respect to [1$\bar{1}$0], [110]
by an angle different from multiples of $90^{\circ}$ the
structured sample lacks any symmetry operations except for the
identity and corresponds to the point group C$_1$.
Phenomenologically, in this case each of the photocurrents
component $j_{x}, j_{y}$ is a sum of a
polarization-independent term and three terms proportional to
$|e_{x}|^2 - |e_{y}|^2, \{ e_{x} e^*_{y}\}$ and $P_c$, and
the normal-incidence PGE is described by eight independent
coefficients.

Equations (\ref{2}) describe the normal-incident photocurrent for
any in-plane orientation of the axes $x, y$ if the microscopic
asymmetry of an unstructured quantum well is ignored and only
asymmetry of the lateral superlattice is taken into account. If
the pair of contacts makes an angle of 45$^{\circ}$ with the axes
$x$ and $y$ the normal-incidence photocurrent is an equal
superposition of the currents $j_{x}$ and $j_{y}$. If the initial
laser light is polarized along the line $l$ connecting the
contacts then, for normal incidence, the dependence of the
photocurrent component $j_l$ along this line on the orientation of
the $\lambda/4$ and $\lambda/2$ plates is given by
\begin{equation} \label{3phi}
j_l( \varphi)/I = \gamma_l \sin{2 \varphi} + \frac{\chi_{3l}}{2} \sin{4 \varphi} +
\frac{\chi_{2l}}{2} \cos{4 \varphi} + \chi_{1l} + \frac{\chi_{2l}}{2}\:,
\end{equation}
and
\begin{equation} \label{3alpha}
 j_l(\alpha)/I = \chi_{3l} \sin{2 \alpha} + \chi_{2l} \cos{2 \alpha} + \chi_{1l}\:,
\end{equation}
respectively.  These equations agree with the experimental
polarization dependences of the current  $J \propto
j_l$ [see Figs.~\ref{fig2}, \ref{fig4} and Eqs.~(\ref{phenom1}),
(\ref{phenom2})] yielding
$$
\gamma_l = a, \,\,\, \,\,\, \,\,\, \frac{\chi_{3l}}{2} = b,  \,\,\,\,\,\, \,\,\, \frac{\chi_{2l}}{2} = c,
\,\,\,\,\,\, \,\,\, \chi_{1l} = d - c \:.
$$
The dependence of the photocurrent $j_l$ on the angle of incidence
$\theta_0$ can be described by the factor $ \cos{\theta_0} \xi$.
This also agrees with experimental observations which are well
fitted by Eqs.~(\ref{jd}) and (\ref{jrefcirc}), see
Fig.~\ref{fig3} which shows the CPGE and LPGE contributions given
by coefficients $a$ and $d$, respectively.

An additional mechanism of the asymmetry leading to the
photocurrents can be related to the space-modulated intensity of
the radiation exciting the structure. Such an inhomogeneous
distribution of the electric field in structured samples with
distance between the groove edges and QW layers of nanometer scale
is expected due to near-field effects. In the THz range a local
enhancement of electric fields in structures with subwavelength
pattern has previously been observed in GaAs tunnelling Schottky
barrier junctions ( for review see Chapter 2 in the
book~\cite{book}). In our samples, due to the near-field effects,
the amplitude of a plane electromagnetic field penetrating through
the superimposed grating becomes a periodic function of $x$ with
the same period $a$. In the asymmetrical structure the potential
$V(x_c)$ and the intensity $I(x_c)$ of the normally-incident
radiation can be shifted relative to each other in phase. As a
result, the product $I(x_c) (dV/dx_c)$ averaged over  space as
well as the coefficients $\chi_j$ in Eq.~(\ref{2}) are
nonvanishing.

\section{Microscopical model}

If the lateral superstructure is
responsible for the PGE observed at normal incidence then one can
ignore the initial symmetry C$_{2v}$ of the reference
heterostructure, disregard mechanisms of PGE related to the lack
of an inversion center in the unstructured sample and rely only on
the symmetry of superstructure potential $V(x)$ and the in-plane
intensity modulation. In this case one can apply Eqs.~(\ref{2})
for any orientation of the axes $x, y$ irrespectively to the
crystallographic directions [1$\bar{1}$0], [110].

We have analyzed microscopic mechanisms of PGE by using the classical
Boltzmann equation for the electron distribution function $f_{\bm k}$, namely,
\begin{equation} \label{4}
\left( \frac{\partial}{\partial t} + v_{\bm k,x}
\frac{\partial}{\partial x} + \frac{\bm F}{\hbar}
\frac{\partial}{\partial {\bm k}} \right)
f_{\bm k}(x) + Q^{(p)}_{\bm k} + Q^{(\varepsilon)}_k  = 0\:.
\end{equation}
Here the simplified notation $x$ is used for the coordinate $x$,
${\bm k}$ is the electron two-dimensional wave vector, ${\bm F}$
is a sum of the time-dependent electric-field force
\begin{equation} \label{4a}
e {\bm E}(t) = e ({\bm E}_0 {\rm e}^{- {\rm i} \omega t} +
{\bm E}^*_0 {\rm e}^{{\rm i} \omega t})
\end{equation}
of the light wave and the static force $- dV(x)/dx$, $\omega$ is
the light frequency, ${\bm v}_{\bm k} = \hbar {\bm k}/m^*$ is the
electron velocity, $e$ and $m^*$ are the electron charge and
effective mass, $Q^{(p)}_{\bm k}$ and $Q^{(\varepsilon)}_k$ are
the collision terms responsible for the electron momentum and
energy relaxation, respectively. The operator $Q^{(p)}_{\bm k}$ is
taken in the simplest form
\begin{equation} \label{4b}
Q^{(p)}_{\bm k} = \frac{ f_{\bm k} - \langle f_{\bm k} \rangle }{\tau}\:,
\end{equation}
where $\tau$ is the momentum relaxation time and the brackets mean
averaging over the directions of ${\bm k}$. The operator
$Q^{(\varepsilon)}_k$ acts on the distribution function averaged
over the directions of ${\bm k}$ and depends only on the modulus
$k = |{\bm k}|$. Equation (\ref{4}) is valid for the weak and
smooth potential satisfying the conditions $|V(x)| \ll
\varepsilon_e$ and $q_0 \equiv 2 \pi/a \ll k_e$, where $k_e$ and
$\varepsilon_e$ are the typical electron wave vector and energy,
and for the photon energy $\hbar \omega$ being much smaller than
$\varepsilon_e$.

If the space modulation of the radiation intensity is ignored the
photocurrent is obtained by solving Eq.~(\ref{4}) in the fifth
order  perturbation theory, namely, the second
order in the amplitude of the light electric field ${\bm E}_0$ and
the third order in the lateral potential $V(x)$
\begin{equation} \label{5}
j_l = R_l |{\bm E}_0|^2 \overline{\left( \frac{dV}{dx}\right)^3}\equiv R_l
|{\bm E}_0|^2 \zeta (q_0 \tilde{V})^3\:.
\end{equation}
Here $\tilde{V}^2$ is the dispersion $\overline{V(x)^2}$, the
overline means averaging over  space (without losing generality,
we suppose $\overline{V(x)} = 0$), and $\zeta$ is a dimensionless
measure of the potential asymmetry. For the simplest asymmetric
potential $V_1 \cos{(q_0 x)} + V_2 \sin{(2q_0 x)}$ with
$q_0=2\pi/a$, the parameter $\zeta$ equals $- 3 \sqrt{2} V_1^2
V_2/(V_1^2 + V_2^2)^{3/2}$ and $\tilde{V}^2 = (V_1^2 + V_2^2)/2$.
The form of the coefficient $R_l$ in Eq.~(\ref{5})
depends on the radiation polarization, in accordance with
Eq.~(\ref{2}), and the experimental conditions, particularly, on
the relation between the light frequency $\omega$ and the momentum
and energy relaxation times, $\tau$ and $\tau_{\varepsilon}$,
respectively, as well as on the relation between the period $a$
and the electron free-path length $l_e$ and energy diffusion
length $l_{\varepsilon}$.
\section{Estimations for photogalvanic currents}
{\it Circular PGE}. First of all, we present an estimation for the
circular photocurrent described by the coefficient $\gamma$ in the
second equation (\ref{2}) for the non-degenerate electron gas in
the limiting case $\tau_{\varepsilon}^{-1}, q_0 l_e, q_0
l_{\varepsilon} \ll \omega$:
\begin{equation} \label{7}
j_y = \gamma I \approx e \nu g \tau  N \frac{\hbar k_e}{m^*} \left( \frac{q_0}{k_e}\right)^3
\frac{  \zeta \tilde{V}^3 }{k_B T (\hbar \omega)^2 }\:,
\end{equation}
where $\nu = d \ln{\tau}/d \ln{\varepsilon}$, $g$ is the photon
absorption probability rate per particle:
\begin{equation} \label{7a}
g = \frac{4 \pi e^2 }{m^*
c n_{\omega}} \frac{I}{\hbar \omega}
\frac{\tau}{1 + (\omega \tau)^2}\:.
\end{equation}
The radiation intensity $I$ is related to the amplitude ${\bm
E}_0$ by
\[
I = \frac{c n_{\omega}}{2 \pi} (|E_{0x}|^2 + |E_{0y}|^2)\:,
\]
$n_{\omega}$ is the refraction index and $c$ is the light velocity
in vacuum. While deriving Eq.~(\ref{7}) we assumed $g$ to be
$x$-independent.

{\it Polarization-independent PGE}. The photocurrent independent
of polarization and proportional to the coefficient $\chi_1$ in
Eq.~(\ref{2}) can be related to heating of free carriers by the
electromagnetic wave. At high temperatures the conditions $l_e,
l_{\varepsilon} \ll a$ are fulfilled, and the kinetic equation
(\ref{4}) can be reduced to the macroscopic equations for the
two-dimensional electron density $N(x)$, local nonequilibrium
temperature $\Theta(x)$, current density $j_x$ and energy flux
density $i_{\varepsilon,x}(x)$. In the continuity equation for the
energy flux density the supplied and dissipated powers are taken
in the form
\begin{equation} \label{WW}
W_{\varepsilon}^{({\rm in})} = \hbar \omega g N\:,\:W_{\varepsilon}^{({\rm out})} =
\frac{k_B(\Theta - T)}{\tau_{\varepsilon}}\  N\:,
\end{equation}
where $\tau_{\varepsilon}$ is the energy relaxation time, $\Theta$
is the local electron temperature, $T$ is the equilibrium phonon
temperature and $k_B$ is the Boltzmann constant. In what follows
we assume a moderate pump power resulting in a weak increase in
the temperature, $\Theta - T \ll T$. Under the homogeneous optical
excitation the macroscopical equations have the following solution
\[
k_B \Theta = k_B T + \hbar \omega g \tau_{\varepsilon}\:,\: N(x) = N_0 {\rm e}^{- V(x)/k_B \Theta}\:,
\]
where $N_0$ is $x$-independent. For this solution both the
electric current $j_x$ and the flux $i_{\varepsilon,x}$ are
absent. However, a value of $j_x$ becomes nonzero with allowance
for the generation rate $g$ to vary in space. Let this variation
be described by $g(x) = g_0 + g^{(1)} \cos{(q_0 x + \varphi_g)}$.
The steady-state inhomogeneous generation produces a permanent
periodic electron temperature
\[
k_B \Theta(x) = k_B T + \tau_{\varepsilon} \hbar \omega g^{(1)} \cos{(q_0 x + \varphi_g)}
\]
which is followed by a light-induced correction to the
space-oscillating contribution to the electron density
\[
\delta N(x) = - \frac{N_0 \tau_{\varepsilon}}{k_B T} \hbar \omega g^{(1)} \cos{(q_0 x + \varphi_g)}\:.
\]
The photocurrent is calculated as an average
\[
j_x = - \mu \overline{\frac{d V(x)}{dx} \delta N(x)}\:,
\]
where $u$ is the mobility $e\tau/m^*$. For the lateral potential
taken in the form $V(x) = V_1 \cos{(q_0 x + \varphi_V})$, the
symmetry of the system is broken due to a phase shift between
$V(x)$ and $\Theta(x)$. The result for $\chi_1$ reads
\begin{eqnarray} \label{chi1}
j_x = \chi_1 I &=& \mu N_0  q_0 V_1  \frac{\hbar \omega g^{(1)} \tau_{\varepsilon}}{2k_B T}\
\sin{(\varphi_g - \varphi_V)} \\&=& \mu N_0 \hbar q_0 \zeta' g_0 \frac{V_1}{2k_B T}
\omega \tau_{\varepsilon}
\:. \nonumber
\end{eqnarray}
Here $\zeta' = (g^{(1)}/g_0)\sin{(\varphi_g - \varphi_V)}$ is the
parameter of asymmetry related to the inhomogeneous
photoexcitation. The model used to derive Eq.~(\ref{chi1}) is
similar to the model considered in Refs.~\cite{buttiker,buttiker2}
for a ratchet with sinusoidal potential and temperature and a
relative phase lag between them. The difference is that here we
assume $V_1 \ll k_B T$ while in Ref.~\cite{buttiker2} the opposite
case is considered. It should be noted that the nonequilibrium
asymmetric systems with both a periodic potential $V(x)$ and a
periodic temperature profile $T(x)$ are referred to as the Seebeck
ratchets~\cite{reimann}.

It follows from Eqs.~(\ref{7a}) and (\ref{chi1}) that
microscopically the coefficient $\chi_1$ can be presented in the
form
\begin{equation} \label{chi1mic}
\chi_1 = \frac{2 \pi e^2 }{\hbar c n_{\omega} } \zeta' \mu N_0 \frac{\hbar q_0}{m^*}
\frac{\tau \tau_{\varepsilon}}{1 + (\omega \tau)^2} \frac{V_1}{k_B T}\:.
\end{equation}

{\it Circular PGE under inhomogeneous excitation}. Now we show
that the polarization-dependent photocurrents and even the CPGE
currents can as well be induced in a lateral superlattice with the
out-of-phase periodic potential $V(x)$ and generation $g(x)$. The
photocurrent in the $y$ direction is calculated according to
\begin{equation} \label{jy}
j_y = \frac{2 e^2 \tau}{m^*} {\rm Re} \{\overline{E_{0y}^*(x) \delta N_{\omega}(x)} \}
\:,
\end{equation}
where $\delta N_{\omega}(x)$ is the amplitude of the electron
density oscillation linear in the THz electric field $E_{0x}$.
From the continuity equation $- {\rm i} \omega e \delta N_{\omega}
+ dj_{x,\omega}/dx = 0$ and the equation for the linear-response
electric current contribution modulated in space
\[
j_{x,\omega} = - \frac{e^2 \tau N_0}{m^*} \frac{E_{0x}}{1 - {\rm i} \omega \tau} \frac{V(x)}{k_B T}
\]
we find the amplitude $\delta N_{\omega}(x)$ and, finally, the
circular photocurrent
\begin{equation} \label{circ2}
\gamma I = \mu N_0 \hbar q_0 \zeta' g_0  \frac{V_1}{4k_B T}
\:.
\end{equation}
The coefficient $\gamma$ is given by
\begin{equation} \label{gammamic}
\gamma = \frac{\pi e^2 }{\hbar c n_{\omega} } \zeta' \mu N_0 \frac{\hbar q_0}{m^*}
\frac{\tau}{\omega (1 + \omega^2 \tau^2)} \frac{V_1}{k_B T}\:.
\end{equation}

The generation of a steady-state electron flow along the $y$ axis
sensitive to the circular polarization $P_c$ results from two
phase shifts of the oscillation $\delta N(x,t)$, in space with
respect to $V(x)$ by $\varphi_g - \varphi_V$ and in time with
respect to $E_{x}(t)$ by $\arctan{(\omega \tau)}$. This current is
smaller as compared with the polarization-independent current by a
factor of $2 \omega \tau_{\varepsilon}$. This agrees with the
experiment where the polarization-independent contribution is
dominating. The ratio of the circular photocurrents (\ref{7}) and
(\ref{circ2}) equals $(\zeta/\zeta')( q_0 \tilde{V}/k_e \hbar
\omega)^2 $. Due to the small parameter $(q_0/k_e)^2$ the
contribution (\ref{circ2}) is expected to exceed the alternative
contribution (\ref{7}). For the inhomogeneous photoexcitation, the
coefficients $\chi_2$ and $\chi_3$ have the same order of
magnitude  $ \sim\gamma \omega \tau$.

\section{Summary} A new mechanism of circular and linear photogalvanic
effects has been proposed and demonstrated experimentally. The
lateral grating etched into the sample's surface induces a
periodical lateral potential acting on the two-dimensional
electron gas. In addition, it modifies the normally-incident
radiation causing its spatial modulation in plane of the electron
gas. If the lateral structuring is asymmetrical the spatial
modulations of the static lateral potential $V(x)$ and the
radiation intensity $I(x)$ are relatively shifted relatively to
each other. As a result the product of the static force $-
dV(x)/dx$ and the photothermal modulation of the electron density
$\delta N(x)$ has a nonzero space average and, therefore, a
homogeneous electric current is generated, an effect previously
predicted by Blanter and B\"uttiker~\cite{buttiker2}. In this
paper we have proposed polarization-dependent photocurrents
arising in the same system with broken symmetry due to the phase
shift between periodic potential and periodic light field. These
currents, in contrast to the polarization-independent
photocurrent, are independent of the energy relaxation time, if
$\omega \tau_{\varepsilon} \gg 1$, and controlled only by momentum
relaxation time.

\acknowledgments The financial support from the DFG and RFBR is
gratefully acknowledged. E.L.I. thanks DFG for the Merkator
professorship. We are grateful M.M. Voronov, V.V. Bel'kov, L.E.
Golub and S.A. Tarasenko for fruitful discussions.

\end{document}